\documentclass[paper]{JHEP3} % 10pt is ignored!

\JHEPspecialurl{http://jhep.sissa.it/JOURNAL/JHEP3.tar.gz}

\usepackage{epsfig}
\usepackage{cite}
\usepackage{wick}
\preprint{
KEK-TH-973\\
hep-th/0410263\\}

\title{Perturbative dynamics of fuzzy spheres at large $N$}
\author{Takehiro Azuma, 
Keiichi Nagao, Jun Nishimura \\
Institute of Particle and Nuclear Studies, \\ 
High Energy Accelerator Research Organization (KEK),\\
1-1 Oho, Tsukuba 305-0801, Japan  \\
\email{azumat@post.kek.jp, 
nagao@post.kek.jp, 
jnishi@post.kek.jp}} 

\abstract{
We clarify some peculiar aspects of
the perturbative expansion around a classical fuzzy-sphere solution 
in matrix models with a cubic term.
While the effective action in the large-$N$ limit
is saturated at the one-loop level,
we find that the ``one-loop dominance'' does not hold 
for generic observables due to one-particle reducible diagrams.
However, we may exploit the one-loop dominance for the 
effective action and obtain various observables to all orders 
from one-loop calculation by simply
shifting the center of expansion to the ``quantum solution'', 
which extremizes the effective action.
We confirm the validity of this method
by comparison with the direct two-loop calculation
and with Monte Carlo results
in the 3d Yang-Mills-Chern-Simons matrix model.
From the all order result we find that the perturbative expansion
has a finite radius of convergence.
}

\keywords{Matrix Models, Non-Commutative Geometry}

\newcommand{\bel}{\begin{equation}\label}

\newcommand {\beq}{\begin{equation}}
\newcommand {\eeq}{\end{equation}}
\newcommand {\beqa}{\begin{eqnarray}}
\newcommand {\eeqa}{\end{eqnarray}}
\newcommand {\bc}{\begin{center}}
\newcommand {\ec}{\end{center}}
\newcommand {\tr}{{\rm tr}\,}

\newcommand {\del}{\partial}

\newcommand {\defeq}{\stackrel{\rm def}{=}}

\newcommand {\stwo}{{\rm S}^{2}}

\newcommand {\A}{ {\tilde A}}
\def\dag{\dagger}

\def\vs5{\vspace*{5mm}}
\def\vs1{\vspace*{1cm}}
\def\vs2{\vspace*{2cm}}
\def\hs5{\vspace*{5mm}}
\def\hs1{\hspace*{1cm}}
\def\hs2{\hspace*{2cm}}
\def\vs50{\vspace*{50mm}}
\def\vs20{\vspace*{20mm}}

\def\tr{\hbox{tr}}

\begin{document}

\section{Introduction}
Fuzzy spheres \cite{Madore}, which are simple compact noncommutative 
manifolds, have been studied extensively.
There are various motivations for studying the fuzzy spheres.
First it is expected that
the noncommutative geometry provides a crucial link to string theory and 
quantum gravity.
Indeed the Yang-Mills theory on noncommutative geometry
is shown to emerge from a certain low-energy limit of 
string theory \cite{Seiberg:1999vs}.
There is also an independent observation that the space-time 
uncertainty relation, which is naturally realized by noncommutative
geometry, can be derived from some general assumptions
on the underlying theory of quantum gravity \cite{gravity}.
One may also use fuzzy spheres
as a regularization scheme alternative to 
the lattice regularization \cite{Grosse:1995ar}.
Unlike the lattice, fuzzy spheres preserve the continuous symmetries 
of the space-time considered, and
the well-known problem of chiral symmetry 
\cite{Grosse:1994ed,%
Carow-Watamura:1996wg, chiral_anomaly, non_chi,balagovi,% 
chiral_anomaly2,balaGW,Nishimura:2001dq,AIN,AIN2,Ydri:2002nt,% 
Iso:2002jc,Balachandran:2003ay,nagaolat03, AIN3} 
and supersymmetry in lattice theories may become easier to overcome.

As expected from 
the so-called Myers effect \cite{Myers:1999ps} in string theory,
fuzzy spheres appear as classical solutions in matrix models with
a Chern-Simons term \cite{0101102,0204256,0207115,0301055}.
The properties of the fuzzy spheres in matrix models 
have been studied in refs.\ \cite{0108002,Valtancoli:2002rx,0303120,0309082,%
0312241,0402044,0403242,0405201}. 
One can actually use matrix models to define a regularized 
field theory on fuzzy spheres. Such an approach has been successful
in the case of noncommutative torus \cite{AMNS}, 
where nonperturbative studies have produced 
various important results \cite{Nishimura:2001dq,simNC,Iso:2002jc}. 
These matrix models belong to the class of
so-called large-$N$ reduced models,
which are believed to provide a constructive definition of 
superstring and M theories \cite{9610043,9612115,9703030}.
For instance the IIB matrix model\cite{9612115},
which can be obtained by dimensional reduction of ten-dimensional
${\cal N}=1$ super-Yang-Mills theory, is proposed as 
a constructive definition of type IIB superstring theory.
In this model the space-time is represented by the eigenvalues 
of bosonic matrices, and hence treated as a dynamical object.
The dynamical generation of four-dimensional space-time 
has been discussed in refs.\ \cite{Aoki:1998vn,Ambjorn:2000bf,%
Ambjorn:2000dx,NV,Burda:2000mn,%
Ambjorn:2001xs,exact,sign,Nishimura:2001sx,%
KKKMS,Kawai:2002ub,Vernizzi:2002mu,0307007,Nishimura:2003rj}.

In ref. \cite{0401038} we have performed a first nonperturbative study
on the dynamical properties of the fuzzy spheres, which appear in 
a simple matrix model.
The model can be obtained by dimensional reduction
of 3d Yang-Mills-Chern-Simons (YMCS) theory,
and it incorporates various fuzzy $\stwo$ solutions.
The most important non-perturbative result was
that there exists a first-order phase transition 
as we vary the coefficient of the Chern-Simons term ($\alpha$).
In the small-$\alpha$ phase, the
effect of the Chern-Simons term is negligible and the model behaves 
almost as the pure Yang-Mills model ($\alpha=0$) \cite{9811220}. 
In the large-$\alpha$ phase, on the other hand, 
a single fuzzy sphere appears dynamically
\footnote{
This work has been extended to matrix models 
which incorporate four-dimensional fuzzy manifolds 
as classical solutions \cite{0405096,0405277,0408xxx-s2s2}.
While the fuzzy S$^4$ turned out to be always unstable,
the fuzzy CP$^2$ and the fuzzy ${\rm S}^2 \times {\rm S}^2$
can be stabilized in the large-$N$ limit.
}.
%In this paper we clarify some peculiar properties of the
%perturbative expansion around the fuzzy sphere like solutions.
%For simplicity we consider the 3d YMCS matrix model.
In this phase 
Monte Carlo data agree very well with the one-loop results,
which led us to speculate that the one-loop contribution
dominates the quantum correction
in the large-$N$ limit \cite{0401038}.

In the present paper we first address this issue by direct two-loop
calculation.
While the ``one-loop dominance'' does hold
for the effective action \cite{0303120,0403242,0405201},
we find that this is not the case for generic observables.
The higher-loop contribution that survives the large-$N$ limit actually
comes from one-particle reducible diagrams, which do not appear
in the calculation of the effective action.
However, we may exploit the one-loop dominance for the 
effective action and obtain various observables {\em to all orders} 
from one-loop calculation by simply
shifting the center of expansion to the ``quantum solution'', 
which extremizes the effective action.
%evaluating the one-loop effective action at its extremum 
%\cite{0403242}.
%Indeed we find that the ``all order'' result nicely reproduce
%the behavior of the Monte Carlo data near the critical point.
We confirm the validity of this method
proposed by Kitazawa et al.\ \cite{0403242}
by comparison with the direct two-loop calculation
and with Monte Carlo results.

From the all order result we find that the perturbative expansion
has a finite radius of convergence, and the lower critical point
of the first-order phase transition lies precisely on the
convergence circle. We also reconsider the issue of
the dynamical gauge group, which was previously discussed 
at the one-loop level \cite{0401038}.
The all order calculation of the free energy for
$k$ coinciding fuzzy spheres allows us to obtain a more definite
conclusion on this issue.

This paper is organized as follows. In section \ref{section:model}
we define the model and briefly review the results obtained in 
our previous work.
In section \ref{section:higher_loop}
we perform explicit two-loop calculation of an observable
and demonstrate
that the two-loop contribution survives the large-$N$ limit.
In section \ref{section:all-order-calculations}
we obtain an all order result from one-loop calculation
by shifting the center of expansion to the ``quantum solution''.
%In particular we reproduce the two-loop result obtained
%by the direct calculation in section \ref{section:higher_loop}.
In section \ref{section:other} we extend the all order calculation
to more general observables.
In section \ref{section:MC} we compare the results 
of the all order calculation with
our Monte Carlo results.
In section \ref{section:gauge} we address the issue of
the dynamical gauge group using the all order result
for the free energy. 
%% We conclude that the 
%% gauge group of rank one is generated even if we take into account
%% higher order contributions to the free energy.
Section \ref{section:summary}
is devoted to a summary and concluding remarks.
In appendix \ref{section:diagram_calculation} we give the details
of the two-loop calculation.

\section{Brief review of the model}
\label{section:model}

The model we study in this paper is given by 
the action
 \begin{eqnarray}
  S[A] = N \, \tr \left( - \frac{1}{4} \,  [A_{\mu}, A_{\nu}]^{2}
  + \frac{2}{3} \, i \, \alpha \, \epsilon_{\mu\nu\rho}
  A_{\mu} A_{\nu} A_{\rho}
  \right) \ , \label{model-def}
 \end{eqnarray}
 where $A_{\mu}$ ($\mu=1,2,3$) are $N \times N$ traceless hermitian matrices.
 Here and henceforth we assume that summation is taken
 over repeated Greek indices.
 The classical equation of motion is given by
  \begin{eqnarray}
   [A_{\mu}, [A_{\mu}, A_{\nu}]] + 
   i \, \alpha \, \epsilon_{\nu \sigma \rho}
   [A_{\sigma}, A_{\rho}] = 0 \ .  
  \label{eom}
  \end{eqnarray} 
 As a solution, we consider
  \begin{eqnarray}
   A_{\mu} = X_\mu \defeq \alpha \, L_{\mu} \ , 
  \label{fuzzy-sphere}
  \end{eqnarray}
where $L_{\mu}$ is the $N$-dimensional irreducible
representation of the SU$(2)$ Lie algebra
  \begin{eqnarray}
[L_{\mu}, L_{\nu}] = i \, \epsilon_{\mu \nu \rho} \, L_{\rho} \ .
  \end{eqnarray}
This solution corresponds to the fuzzy ${\rm S}^2$,
and since
 \begin{eqnarray}
  (X_{\mu})^{2} = \frac{1}{4} \, \alpha^{2} \, (N^{2}-1) \, 
 {\bf 1}_{N} \ ,
 \label{casimir}
 \end{eqnarray}
the radius of the sphere is given by $R=\frac{1}{2} 
\alpha \sqrt{N^{2}-1}$.

In ref.\ \cite{0401038} we found that 
the system undergoes a first-order phase transition 
as we vary $\alpha$.
In the large-$\alpha$ regime the dominant configurations
are close to the fuzzy sphere (\ref{fuzzy-sphere}),
while in the small-$\alpha$ regime the large-$N$ property 
is similar to the pure Yang-Mills model ($\alpha = 0$) \cite{9811220},
and the geometry of the dominant configurations is given by 
that of a solid ball.
The two phases are called the ``fuzzy sphere phase'' and the
``Yang-Mills phase'', respectively.
When we discuss the large-$N$ limit in the fuzzy sphere phase,
the natural parameter to fix turned out to be
 \begin{eqnarray}
 {\tilde \alpha} = \alpha \, \sqrt{N} \ .
 \end{eqnarray}
The lower critical point obtained from Monte Carlo simulation is
 \begin{eqnarray}
  {\tilde \alpha}_{\rm cr} \simeq 2.1 \ , 
  \label{critical-mc}
 \end{eqnarray}
which was reproduced later
from the one-loop effective action as
\footnote{This analytical result was informed to us by D. O'Connor
after J.N.\ gave a seminar on the Monte Carlo results
% of ref.\ \cite{0401038} 
including (\ref{critical-mc}) at 
the Dublin Institute for Advanced Studies (DIAS).
Its derivation in section \ref{section:all-order-calculations}
is due to Y.~Kitazawa (private communication).
} 
  \begin{eqnarray}
   {\tilde \alpha}_{\rm cr} = 
\sqrt[4]{\frac{512}{27}}
%\left( \frac{8}{3} \right) ^{\frac{3}{4}}
 = 2.0867794 \cdots \ .
  \label{critical-oneloop}
  \end{eqnarray}
In the fuzzy sphere phase, 
various observables agree well with the one-loop calculation.
We therefore speculated that the one-loop dominance, 
which was previously claimed for the effective action \cite{0303120}, 
holds also for observables.
%This, however, turned out to be not the case as we see in what follows. 
 
\section{Explicit two-loop calculation}
\label{section:higher_loop}

In order to see whether the one-loop dominance
holds also for observables,
we perform explicit two-loop calculation
around the fuzzy sphere solution. 
 We decompose $A_{\mu}$ into the classical background $X_{\mu} = \alpha
L_{\mu}$ and the fluctuation ${\tilde A}_{\mu}$ as
 \begin{eqnarray}
   A_{\mu} = X_{\mu} + \A_{\mu} \ . \label{fs2expansion}
 \end{eqnarray}
  We introduce the following gauge fixing term and the corresponding
  ghost term
  \begin{eqnarray}
   S_{{\rm g.f.}} &=& - \frac{1}{2} \, N \, 
 \tr  \Bigl( [X_{\mu}, A_{\mu}]^{2} \Bigr) \ ,\\
   S_{{\rm gh}} &=& -N \, \tr \, ([X_{\mu}, {\bar c}][A_{\mu}, c]) \ .
  \end{eqnarray}
  The total action can be written as
  \begin{eqnarray}
   S_{{\rm total}} &=& S + S_{{\rm g.f.}} + S_{{\rm gh}} \\
    ~              &=& S[X] + S_{{\rm kin}} + S_{{\rm int}} \ ,
   \label{tot-action}
  \end{eqnarray}
 where the kinetic term $S_{{\rm kin}}$ and the interaction term
 $S_{{\rm int}}$ are given by
  \begin{eqnarray}
  S_{{\rm kin}} &=&   \frac{1}{2}\, N \,  \tr \, ( \A_{\mu} [X_{\lambda},
   [X_{\lambda}, \A_{\mu}]] )
   +  N \, \tr \, ( {\bar c} [X_{\lambda}, [X_{\lambda}, c]] ) \ ,
  \label{kinetic} \\
  S_{{\rm int}} &=& - \frac{1}{4} \, N \, \tr \, ([\A_{\mu}, \A_{\nu}]^{2})
  -N \, \tr \, ( [\A_{\mu}, \A_{\nu}][X_{\mu}, \A_{\nu}]) \nonumber \\
   & & +  \frac{2}{3}\,  i \, \alpha \, N \, \epsilon_{\mu \nu \rho} \, 
        \tr \, ( \A_{\mu} \A_{\nu} \A_{\rho}) 
   -  N \, \tr \, ([X_{\mu}, {\bar c}] [\A_{\mu},c] ) \ .
  \label{interaction}
  \end{eqnarray}

 The free energy $W$ defined by
 \begin{eqnarray}
  e^{-W} = \int d \A \, dc \, d{\bar c} \,
  e^{-S_{\rm total}} \ 
  \label{w-eff}
 \end{eqnarray}
 can be calculated as a perturbative expansion 
 \begin{eqnarray}
  W = \sum_{j=0}^{\infty} W_{j} \ ,
   \label{w-eff-perturbation}
 \end{eqnarray}
 where $W_{j}$ represents the $j$-th order contribution.
%  is proportional to $\alpha^{4(1-j)}$. 
 The first two terms are obtained as \cite{0101102,0401038}
 \begin{eqnarray}
 \label{w-eff-tree}
  W_{0} &=& S[X] = - \frac{{\tilde \alpha}^{4}}{24} (N^{2}-1) \ , \\
  W_{1} &=& \frac{1}{2} \sum_{l=1}^{N-1} (2l +1) \log
  \Bigl[ {\tilde \alpha}^{2} \, l \, (l+1) \Bigr] \ . 
  \label{w-eff-one-loop2}
 \end{eqnarray}
 In order to calculate $W_2$, we have to evaluate the two-loop diagrams
\footnote{
 The diagrams (a)$\sim$(d) are the same as the ones that appear
 in ref.\ \cite{0303120}.
 The  diagram (e) of ref.\ \cite{0303120}, which involves
 a fermion loop, does not appear in the present bosonic model.}
 depicted in figure \ref{twoloop-feyn}.
 The solid line and the dashed line represent the propagators of 
 $\tilde{A}_\mu$ and the ghost, respectively.
 The three-point vertices with (without) a dot represent 
 the third (second) term in (\ref{interaction}).

     \FIGURE{
    \epsfig{file=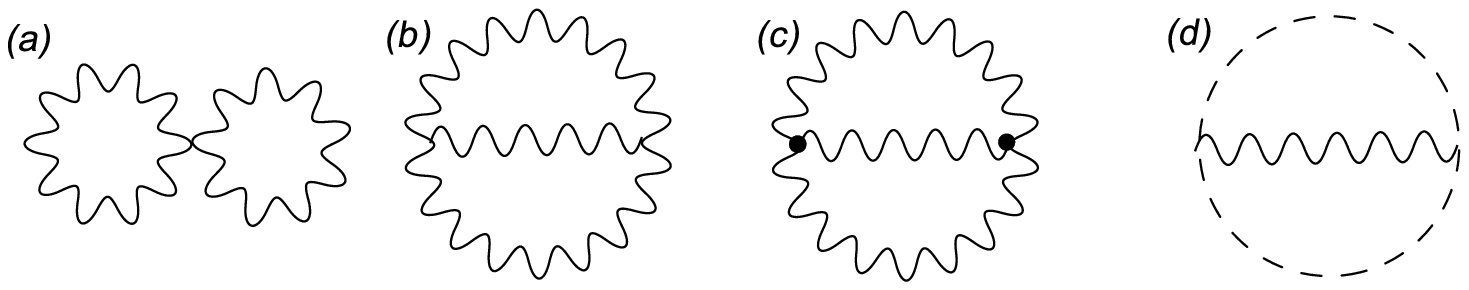,height=2.5cm}
    \epsfig{file=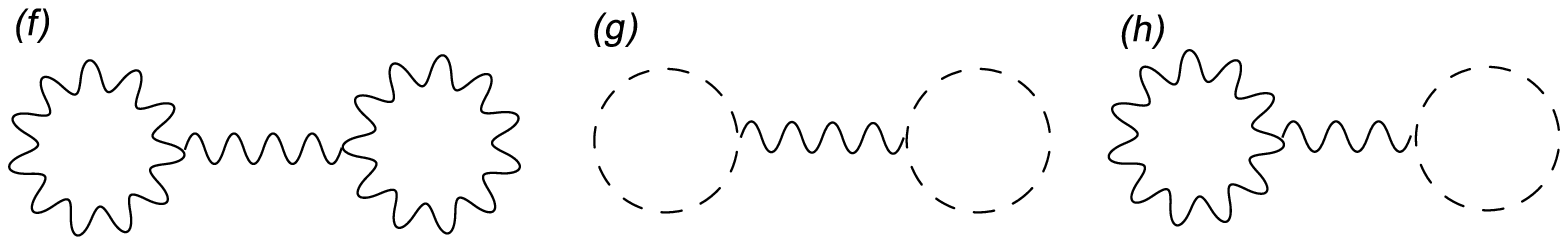,height=2cm}
   \caption{The two-loop diagrams that appear in the perturbative
 calculation of the free energy.
 The diagrams (a) $\sim$ (d) are one-particle irreducible, 
 while the rest are not.}
   \label{twoloop-feyn}}

 As a fundamental observable, let us consider the vacuum expectation
 value of the action $\langle S \rangle$. 
 The perturbative expansion of this quantity can be readily obtained
 from the results for the free energy (\ref{w-eff-perturbation}).
 Let us define the rescaled action
  \begin{eqnarray}
    S(\lambda, \alpha) = \lambda \, N 
   \tr \left( - \frac{1}{4} \,  [A_{\mu}, A_{\nu}]^{2}
  + \frac{2}{3} \, i \, \alpha \, \epsilon_{\mu\nu\rho}
  A_{\mu} A_{\nu} A_{\rho}
  \right) \ , 
  \end{eqnarray}
 and the corresponding free energy
  \begin{eqnarray}
   e^{-W(\lambda, \alpha)} = \int dA \, dc \, d{\bar c} \, 
  e^{-S(\lambda, \alpha)} \ ,
  \end{eqnarray}
 which is related to the original free energy 
 $W=W(1,\alpha)$ through
  \begin{eqnarray}
   W(\lambda, \alpha) = \frac{3}{4} \, (N^{2}-1) \, \log \lambda
  + W(1,\lambda^{\frac{1}{4}}\alpha  ) \ .
  \end{eqnarray}
 Then the observable $\langle S \rangle$ can be written as
  \beqa
  \frac{1}{N^{2}} \langle S \rangle &=& \frac{1}{N^{2}}
  \left. \frac{\partial W(\lambda,\alpha)}{\partial
   \lambda} \right|_{\lambda=1}   \nonumber \\
  &=&    \frac{3}{4} \left(1 - \frac{1}{N^{2}} \right)
     + \frac{\tilde \alpha}{4 N^{2}} 
   \frac{\partial W}{\partial {\tilde \alpha}} \  .
 \label{s-res}
  \eeqa
  Using eqs.\ (\ref{w-eff-tree}) and (\ref{w-eff-one-loop2}),
  we obtain the one-loop result
  \begin{eqnarray}
  \frac{1}{N^{2}} \langle S \rangle_{\rm 1-loop}
   = \left( -\frac{1}{24} {\tilde \alpha}^{4} +1 \right)
      \left( 1 - \frac{1}{N^{2}} \right) \ .
 \label{s-1-loop}
  \end{eqnarray}

 The effective action can be obtained by restricting the diagrams that
 appear in the perturbative expansion of the free energy to one-particle
 irreducible (1PI) diagrams.
 It is the same as the free energy 
 at the one-loop level, but starts to deviate at the two-loop level.
 In ref.\ \cite{0303120} the two-loop effective action was calculated 
 by evaluating the 1PI diagrams (a)$\sim$(d) in figure \ref{twoloop-feyn},
 and the two-loop effect turned out to vanish in the large-$N$ limit.
 However, in order to calculate the observable $\langle S \rangle$, we
 need to evaluate the one-particle reducible (1PR) 
 diagrams (f)$\sim$(h) in figure 
 \ref{twoloop-feyn}, which actually turn out to survive the large-$N$
 limit.

 If we parametrize the higher order contributions
 $W_j$ ($j \ge 2$) to the free energy in eq.\ (\ref{w-eff-perturbation})
  as
 \begin{eqnarray}
  W_j  =  - N^{2} \,  w_{j}(N) \, 
{\tilde \alpha}^{4(1-j)}
%\frac{{\tilde \alpha}^{4(1-j)}}{j-1}
 \ , \label{w-eff-higher}
 \end{eqnarray}
 the perturbative expansion of the observable $\langle S \rangle$
 can be written as
  \begin{eqnarray}
  \frac{1}{N^{2}} \langle S \rangle
   =   \frac{1}{N^{2}} \langle S \rangle_{\rm 1-loop}
   + \sum_{j=2}^{\infty}  (j-1) w_{j}(N) \, {\tilde \alpha}^{4(1-j)} \ .
 \label{s-res2}
  \end{eqnarray}
 The contribution of the 1PI diagrams to $w_2(N)$ is calculated as
 \cite{0303120}
 \begin{eqnarray}
  w_2^{\rm (1PI)}(N)  =
  \frac{1}{N} \Bigl\{ F_{1}(N) + 4 \, F_{3} (N) \Bigr\} \ ,
 \label{1PIw2}
 \end{eqnarray}
 where the functions $F_{1}(N)$ and $F_{3}(N)$ are defined
 by eqs.\ (\ref{func-f1}) and (\ref{func-f3}).
 The large-$N$ behavior is found to be 
 \begin{eqnarray}
  w_2^{\rm (1PI)}(N) 
   \simeq {\rm O} \left(\frac{(\log N)^{2}}{N^{2}} \right) \ ,
 \end{eqnarray}
which vanishes in the large-$N$ limit.

\FIGURE{
  \epsfig{file=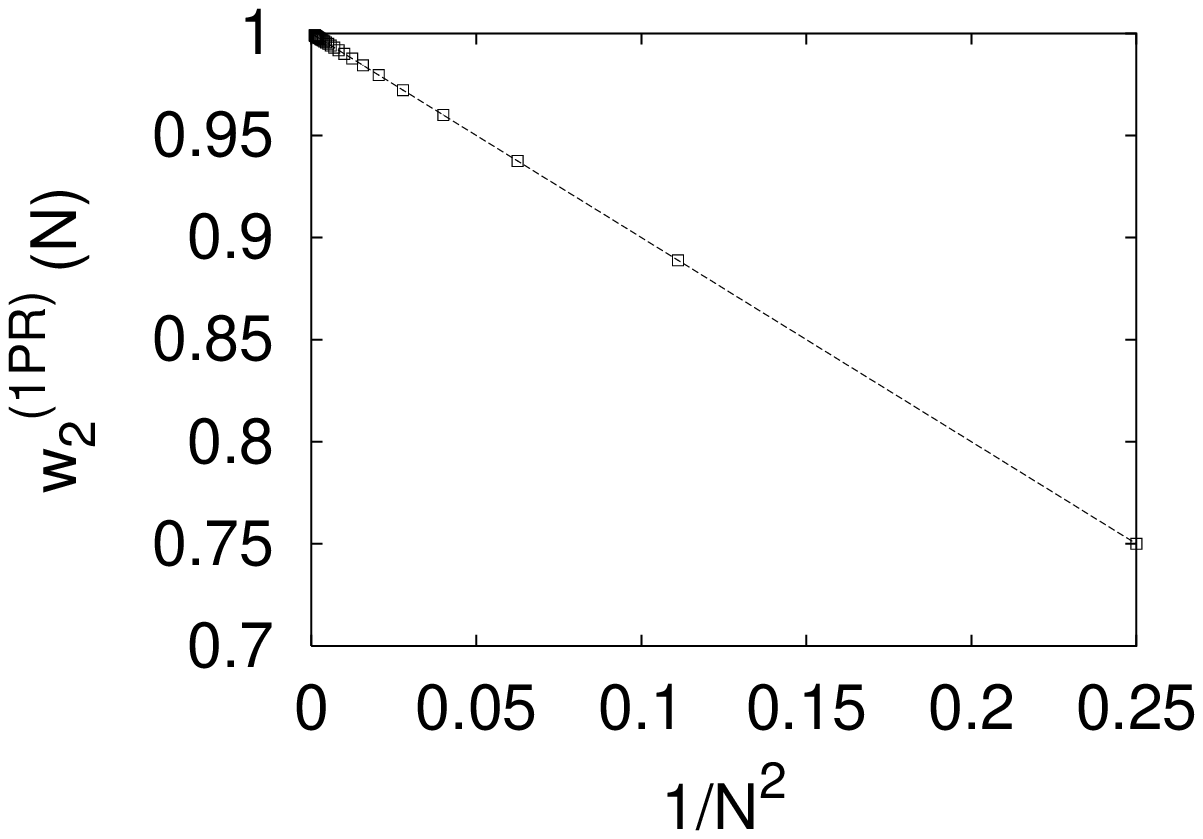,width=7.4cm}
  \caption{The coefficient $  w_2^{\rm (1PR)}(N)$, which comes from
 the 1PR two-loop diagrams, is plotted against $\frac{1}{N^2}$ for
$N=2,3,4,\cdots , 32$.
The straight line represents 
$  w_2^{\rm (1PR)}(N)  =
   1 -  \frac{1}{N^{2}}$.
%is a fit to (\ref{order-diag-g})
%with $a=b=1$.
} \label{diag-g-graph}}

On the other hand,
the contribution of the 1PR diagrams to $w_2(N)$ are calculated as
  \begin{eqnarray}
  w_2^{\rm (1PR)}(N)  =
  \frac{1}{2N} F_{4}(N)  \ ,
  \label{s-diag-h} 
 \end{eqnarray}  
where $F_{4} (N)$ is given explicitly by eq.\ 
(\ref{cubic-cubic-result2}).
In figure \ref{diag-g-graph} we plot $w_2^{\rm (1PR)}(N)$
against $\frac{1}{N^2}$. We find that
  \begin{eqnarray}
  w_2^{\rm (1PR)}(N)  =
   1 -  \frac{1}{N^{2}}
   \label{order-diag-g}
  \end{eqnarray}
within the machine precision for $N=2,3,4,\cdots ,32$.
Thus the 1PR diagrams yield the contribution to $w_2(N)$
which survives the large-$N$ limit.
Summing the contributions from the 1PI diagrams and the 1PR diagrams,
we obtain the large-$N$ behavior
  \begin{eqnarray}
  w_2(N)  =    
  w_2^{\rm (1PI)}(N) +   w_2^{\rm (1PR)}(N) 
  = 1 + {\rm O} \left(\frac{(\log N)^{2}}{N^{2}} \right) \ .
   \label{w2_asymptotics}
  \end{eqnarray}

 \section{All order result from one-loop calculation}
\label{section:all-order-calculations}

In this section we apply a method in ref.\ \cite{0403242}
to obtain an all order result for the observable 
from one-loop calculation.
The crucial point is that the free energy and the effective
action are related to each other by the Legendre transformation.
Therefore, one can obtain the free energy
by evaluating the effective action at its extremum.
Since the effective action enjoys the one-loop dominance
in the case at hand, we may obtain the free energy,
and hence the observable,
to all orders in $\frac{1}{\tilde{\alpha}^4}$ 
in the large-$N$ limit.
%We confirm that the two-loop result obtained in the previous
%section can be correctly reproduced from the {\em one-loop} effective
%action.

By expanding the theory around a rescaled fuzzy-sphere configuration 
$A_{\mu} = \beta \, L_{\mu}$,
we obtain the one-loop effective action $\Gamma$
in the large-$N$ limit as
\footnote{Here and henceforth we neglect an irrelevant constant
term in the effective action and the free energy.}
 \begin{eqnarray}
\lim_{N\rightarrow \infty}
  \frac{1}{N^{2}} \, \Gamma (\tilde \beta)
=
%\simeq 
\left( \frac{1}{8} \, {\tilde \beta}^{4}
 - \frac{1}{6} \, {\tilde \alpha} \, {\tilde \beta}^{3} \right)
 + \log {\tilde \beta}\ , \label{effective-largen}
 \end{eqnarray}
where ${\tilde \beta} = \beta \sqrt{N}$.
The function of $\tilde{\beta}$ on the right-hand side
has a local minimum for $\tilde{\alpha}>
\sqrt[4]{\frac{512}{27}}
%\left( \frac{8}{3} \right) ^{\frac{3}{4}}
$,
from which one obtains the critical point ${\tilde \alpha}_{\rm cr}$
in eq.\ (\ref{critical-oneloop}).
The value of $\tilde{\beta}$ which
gives the local minimum can be obtained by solving a fourth order
algebraic equation, and it can be written explicitly as
 \begin{eqnarray}
  {\tilde \beta} = f (\tilde \alpha) \defeq
\frac{1}{4} \, \tilde \alpha \left(
  1 +  \sqrt{1 + \delta} 
   +  \sqrt{ 2 - \delta + \frac{2}{\sqrt{1+\delta}}} \right) \ , 
    \label{saddle-sol}
 \end{eqnarray}
where
 \begin{eqnarray}
  \delta = 4 \, {\tilde \alpha}^{- \frac{4}{3}}
   \left[ \left( 1 + \sqrt{1 - 
   \frac{512}{27 \, {\tilde \alpha}^{4}}} \right)^{\frac{1}{3}}
        + \left( 1 - \sqrt{1 - 
   \frac{512}{27 \,  {\tilde \alpha}^{4}}} \right)^{\frac{1}{3}} \right] \ .
 \end{eqnarray}
Plugging this solution into the one-loop
effective action (\ref{effective-largen}),
we obtain an all order result for the free energy as
\beq
\lim_{N\rightarrow \infty}
\frac{1}{N^{2}}  W =
%\simeq 
\left( \frac{1}{8} \, f (\tilde \alpha)^{4}
 - \frac{1}{6} \, {\tilde \alpha} \, f (\tilde \alpha)^{3} \right)
 + \log f (\tilde \alpha) \ .
\label{W-all-order}
\eeq
By using (\ref{s-res}),
we can readily obtain an all order result for
the observable $\langle S \rangle$ as
\beq
\lim_{N\rightarrow \infty}
   \frac{1}{N^{2}} \langle S \rangle
=
\frac{3}{4} - 
\frac{1}{24} \, \tilde \alpha \, f (\tilde \alpha)^{3} \ ,
\label{S-all-order}
\eeq
where we have used the fact that $\beta = f (\tilde \alpha)$
extremizes the one-loop effective action (\ref{effective-largen}).

%% Let us check that the all order result obtained above
%% includes the two-loop contribution obtained in section 
%% \ref{section:higher_loop}.

In order to check that the two-loop contribution obtained in section 
\ref{section:higher_loop} can be reproduced correctly,
let us expand the all order results at large $\tilde \alpha$.
First the solution (\ref{saddle-sol}) can be expanded in terms of 
$\frac{1}{\tilde{\alpha}^4}$ as
 \begin{eqnarray}
  f ({\tilde \alpha}) = {\tilde \alpha} 
  \left( 1- \sum_{j=1}^{\infty} c_{j} \, {\tilde \alpha}^{-4j} \right)
  = {\tilde \alpha} 
  \left( 1 - \frac{2}{{\tilde \alpha}^{4}}
  - \frac{12}{{\tilde \alpha}^{8}}
  - \frac{120}{{\tilde \alpha}^{12}}
  - \frac{1456}{{\tilde \alpha}^{16}} - \cdots \right) \ .
 \label{effective-sol}
 \end{eqnarray}
The expansions for the free energy and the observable are obtained 
respectively as
 \beqa
\lim_{N\rightarrow \infty}
\frac{1}{N^{2}}  W &=&
%\simeq 
- \frac{{\tilde \alpha}^{4}}{24} + \log {\tilde \alpha}
      - \frac{1}{{\tilde \alpha}^{4}} 
      - \frac{14}{3 {\tilde \alpha}^{8}} 
   - \frac{110}{3 {\tilde \alpha}^{12}}
      - \frac{364}{{\tilde \alpha}^{16}} - \cdots \ ,
\label{Wexpansion} \\
\lim_{N\rightarrow \infty}
   \frac{1}{N^{2}} \langle S \rangle
&=&
% \simeq 
- \frac{{\tilde \alpha}^{4}}{24} + 1 
   + \frac{1}{{\tilde \alpha}^{4}}
   + \frac{28}{3 {\tilde \alpha}^{8}} 
  + \frac{110}{{\tilde \alpha}^{12}} 
  +   \frac{1456}{{\tilde \alpha}^{16}} + \cdots  \ . 
  \label{s-5loop}
 \eeqa
%The term of the order O$({\tilde \alpha}^{4(1-j)})$ 
%corresponds to the $j$-loop contribution.
The third term, which corresponds to the two-loop
contribution, indeed agrees with the result (\ref{w2_asymptotics}),
which we obtained by the direct calculation.

   \FIGURE{
    \epsfig{file=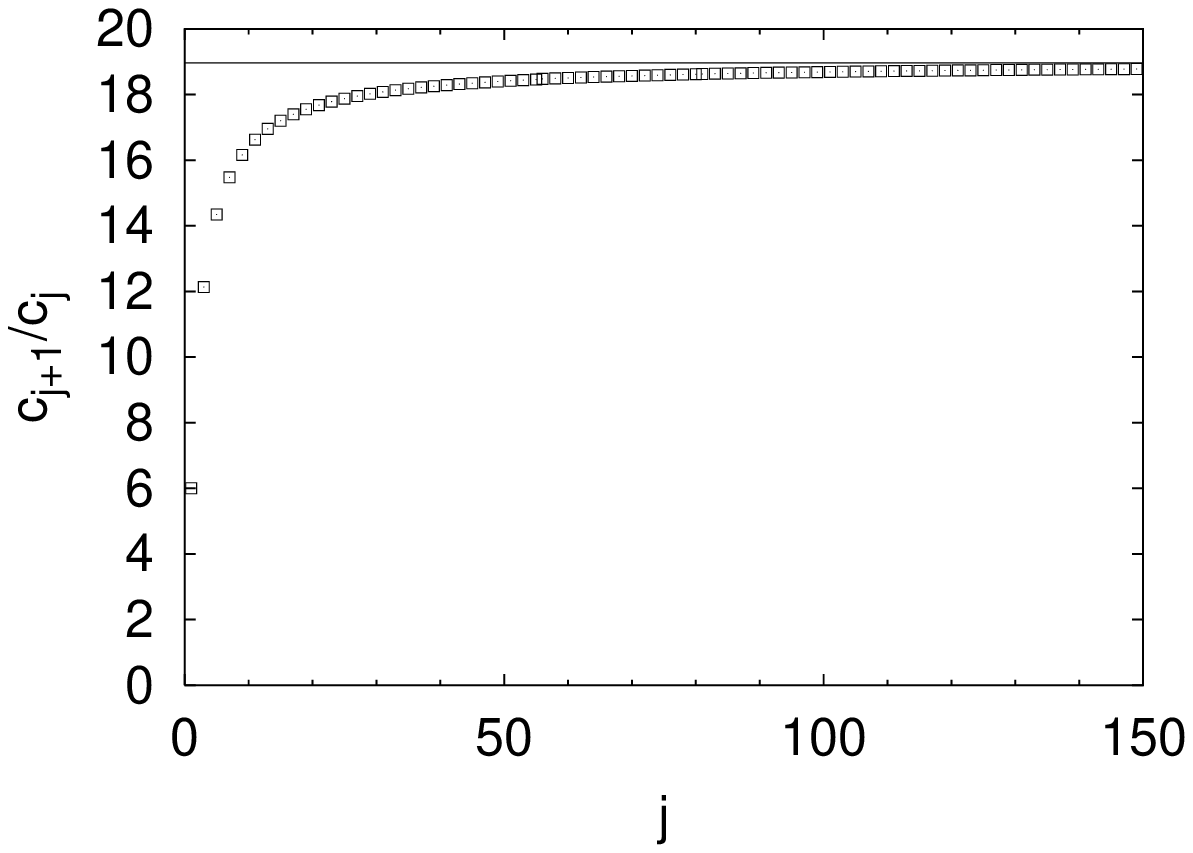,width=7.4cm}
   \caption{The ratio $\frac{c_{j+1}}{c_{j}}$ 
  of the coefficients in (\ref{effective-sol}).
  The horizontal line represents
%  ${\tilde \alpha}_{\rm cr}^{\ 4} = \frac{512}{27}$.
  $ \frac{512}{27} = 18.96296...$.
}
   \label{coef-prop}}

Let us discuss the convergence radius
of these series expansions.
In figure \ref{coef-prop}
we plot the ratio of the coefficients $\frac{c_{j+1}}{c_{j}}$ 
in (\ref{effective-sol}), which is found to converge as
\beq
\lim_{j \to \infty} \frac{c_{j+1}}{c_{j}} 
= \frac{512}{27} = {\tilde \alpha}_{\rm cr}^{\ 4} \ .
\eeq
This implies that the infinite series (\ref{effective-sol}) 
converges for ${\tilde \alpha} > {\tilde \alpha}_{\rm cr}$.
The expansion (\ref{Wexpansion})
for the free energy as well as that for 
the observable (\ref{s-5loop}) has the same property.
Thus in contrast to the perturbation theory in field theories, which 
usually yields merely an asymptotic expansion,
the perturbative expansion around the fuzzy sphere
in the matrix model has a finite radius of convergence.
The lower critical point (\ref{critical-oneloop})
lies precisely on the convergence circle.

%yields a convergent expansion.

\section{All order calculation of other observables} 
\label{section:other}

So far we have focused on a particular observable 
$\langle S \rangle$, which can 
be obtained by differentiating the free energy.
However, the method for deriving all order results is applicable
to general observables \cite{0403242} since the calculation
reduces to the evaluation of the free energy with an appropriate
source term.

Suppose we want to calculate $\langle {\cal O} \rangle$.
Then we consider an action
\beq
S_{\epsilon} = S + \epsilon \,  {\cal O} \ ,
\label{classicalS_eps}
\eeq
and calculate the corresponding free energy $W_{\epsilon}$,
which has the expansion
\beq
W_{\epsilon} = W
+ \epsilon \, \langle {\cal O} \rangle + {\rm O}(\epsilon^2) \ .
\eeq
%% The VEV (vacuum expectation value) $\langle {\cal O} \rangle$
%% is obtained for various observables
%% at the one-loop level \cite{0401038}.

In order to calculate the free energy to all orders, we first 
calculate the effective action for the system (\ref{classicalS_eps})
around the rescaled configuration $A_{\mu} = \beta \, L_{\mu}$.
Then the free energy $W_{\epsilon}$ can be obtained by evaluating the 
effective action at its extremum.
Since the effective action enjoys the one-loop dominance,
we obtain the free energy to all orders from the one-loop effective 
action. Let us expand the one-loop effective action
$\Gamma_{\epsilon}$
% can be expanded 
as
\beq
\Gamma_{\epsilon}(\tilde \beta ) = \Gamma(\tilde \beta )
%+ \epsilon \, \langle\! \langle {\cal O} \rangle \! \rangle
+ \epsilon \, \Gamma_1 (\tilde \beta ) 
%\Gamma^{(1)} (\tilde \beta ) 
+ {\rm O}(\epsilon^2) \ .
\label{Gam-eps}
\eeq
%where $\langle \! \langle {\cal O} \rangle \! \rangle$ 
%represents the sum of the {\em 1PI diagrams} which
%appear in the perturbative evaluation of the VEV of ${\cal O}$.
The ``quantum solution'' is given by solving the equation
\beq
\frac{\del }
{\del \tilde \beta} 
\, \Gamma_{\epsilon}(\tilde \beta)
= 0 \ ,
\eeq
whose solution is denoted as
\beq
\tilde \beta = f(\tilde \alpha) + \epsilon \, g(\tilde \alpha)+ 
 {\rm O}(\epsilon^2) \ .
\eeq
Note that the first term is given by (\ref{saddle-sol}),
and the second term represents a shift due to the source term.
 
By plugging this solution into (\ref{Gam-eps})
and extracting the ${\rm O} (\epsilon)$ term, 
we obtain the ${\rm O} (\epsilon)$ term of the free energy,
which is nothing but $\langle {\cal O} \rangle$.
Thus we obtain
\beq
\langle {\cal O} \rangle 
=  
% \left. \langle\! \langle {\cal O} \rangle\!\rangle 
% \right|_{\tilde \beta = f(\tilde \alpha)} 
\Gamma_{1} \Bigl( f(\tilde \alpha) \Bigr) 
+ g(\tilde \alpha) 
\left. \frac{\del \, \Gamma}{\del \tilde \beta} 
\right|_{\tilde \beta = f(\tilde \alpha)} \ .
\label{obs-all}
\eeq
Note that 
the second term vanishes since
$\tilde \beta = f(\tilde \alpha)$
extremizes $\Gamma (\tilde \beta)$,
and the first term can be obtained by omitting
the 1PR diagrams in the 
one-loop calculation 
of the observable $\langle {\cal O} \rangle$
and replacing $\tilde \alpha$ by $f(\tilde \alpha)$.

As a concrete observable let us consider the spacetime extent 
$\left\langle \frac{1}{N} \tr (A_\mu)^{2} \right\rangle$. 
The one-loop result is given as \cite{0401038}
  \begin{eqnarray}
\lim_{N\rightarrow \infty}
   \frac{1}{N} \left\langle \frac{1}{N} \tr (A_\mu)^{2} 
 \right \rangle_{\rm 1-loop}
   = \frac{1}{4}\, {\tilde \alpha}^{2}
   - \frac{1}{{\tilde \alpha}^{2}} \ . 
  \label{a-sq-1loop}
  \end{eqnarray}
 The first (second) term corresponds to the classical 
 (one-loop) contribution.
 As can be seen from (C.8) of ref.\ \cite{0401038}, 
 the one-loop contribution is given totally by a tadpole diagram,
 which is one-particle reducible.
 Therefore, the all order result can be obtained as
 \begin{eqnarray}
\lim_{N\rightarrow \infty}
  \frac{1}{N} \left\langle \frac{1}{N} \tr (A_\mu)^{2} \right\rangle 
= \frac{1}{4} \, f({\tilde \alpha})^{2}
= 
  \frac{1}{4} \, {\tilde \alpha}^{2}
  - \frac{1}{{\tilde \alpha}^{2}}
   - \frac{5}{{\tilde \alpha}^{6}} - \frac{48}{{\tilde \alpha}^{10}}
   - \frac{572}{{\tilde \alpha}^{14}} - \cdots \ . \label{exact-a-sq}
 \end{eqnarray}

 Next we calculate the Chern-Simons term
 \begin{eqnarray}
 M = \frac{2 \, i}{3\, N} \, \epsilon_{\mu \nu \rho} \, 
 \tr (A_{\mu} A_{\nu} A_{\rho} ) \ .
 \end{eqnarray}
 The one-loop result in the large-$N$ limit
 is given as \cite{0401038}
  \begin{eqnarray}
\lim_{N\rightarrow \infty}
   \frac{1}{\sqrt{N}} \langle M \rangle_{\rm 1-loop} 
   = - \frac{1}{6} \, {\tilde \alpha}^{3}
 + \frac{1}{{\tilde \alpha}} \ ,
   \label{cs-a-1loop}
  \end{eqnarray}
  where the first (second) term corresponds to the classical (one-loop)
  contribution. Similarly to the case of 
  $ \langle \frac{1}{N} \tr (A_{\mu})^{2} \rangle$,
  the one-loop term comes solely from 1PR diagrams.
 Thus we obtain the all order result as
  \begin{eqnarray}
\lim_{N\rightarrow \infty}
   \frac{1}{\sqrt{N}} \langle M \rangle 
 = - \frac{1}{6} \, f({\tilde \alpha})^{3}
 =  - \frac{1}{6} \, {\tilde \alpha}^{3}
   + \frac{1}{\tilde \alpha} + \frac{4}{{\tilde \alpha}^{5}}
   + \frac{112}{3 {\tilde \alpha}^{9}}
   + \frac{440}{{\tilde \alpha}^{13}} + \cdots \ . \label{exact-cs-a}
  \end{eqnarray} 
  
  The all order result for $\langle 
  \frac{1}{N} \tr (F_{\mu \nu})^{2} \rangle$, where
  $F_{\mu \nu} = i \, [A_{\mu}, A_{\nu}]$, can be readily obtained
  by using the exact result \cite{0401038}
%%   \begin{eqnarray}
%%    \left\langle \frac{1}{N} \tr (F_{\mu \nu})^{2} \right\rangle
%%    = 4 \left(  \frac{1}{N^{2}}\langle S \rangle
%%    - \frac{1}{\sqrt{N}} {\tilde \alpha} \langle M \rangle \right) \ . 
%%   \end{eqnarray}  
\begin{eqnarray}
   \left\langle \frac{1}{N} \tr (F_{\mu \nu})^{2} \right\rangle
 + 3 \, \alpha \, \langle M \rangle
   = 3 \left(  1 - \frac{1}{N^2} \right) \ .
  \end{eqnarray}
  Using (\ref{exact-cs-a}), we obtain
  \begin{eqnarray}
 \lim_{N\rightarrow \infty}
   \left\langle \frac{1}{N} \tr (F_{\mu \nu})^{2} \right\rangle
 = 3 + \frac{1}{2} \, {\tilde \alpha} \, f({\tilde \alpha})^{3}
   = \frac{1}{2} \, {\tilde \alpha}^{4}
  - \frac{12}{{\tilde \alpha}^{4}} 
   - \frac{112}{{\tilde \alpha}^{8}}
   - \frac{1320}{{\tilde \alpha}^{12}} - \cdots \ . \label{exact-f-sq}
  \end{eqnarray}

From the relation
  \begin{eqnarray}
\frac{1}{N^{2}}\langle S \rangle
   =   \frac{1}{4} 
       \left\langle \frac{1}{N} \tr (F_{\mu \nu})^{2} \right\rangle
   + \frac{1}{\sqrt{N}}\,  {\tilde \alpha} \, \langle M \rangle  \ , 
  \end{eqnarray}
we find that (\ref{exact-cs-a}) and (\ref{exact-f-sq})
are consistent with the result (\ref{S-all-order}) obtained
in the previous section.
  Clearly the series expansions that appear in this section
  have the same radius of convergence as that for $f(\tilde \alpha)$.
%(\ref{effective-sol}).

\section{Comparison with Monte Carlo results}
\label{section:MC}

  \FIGURE{\epsfig{file=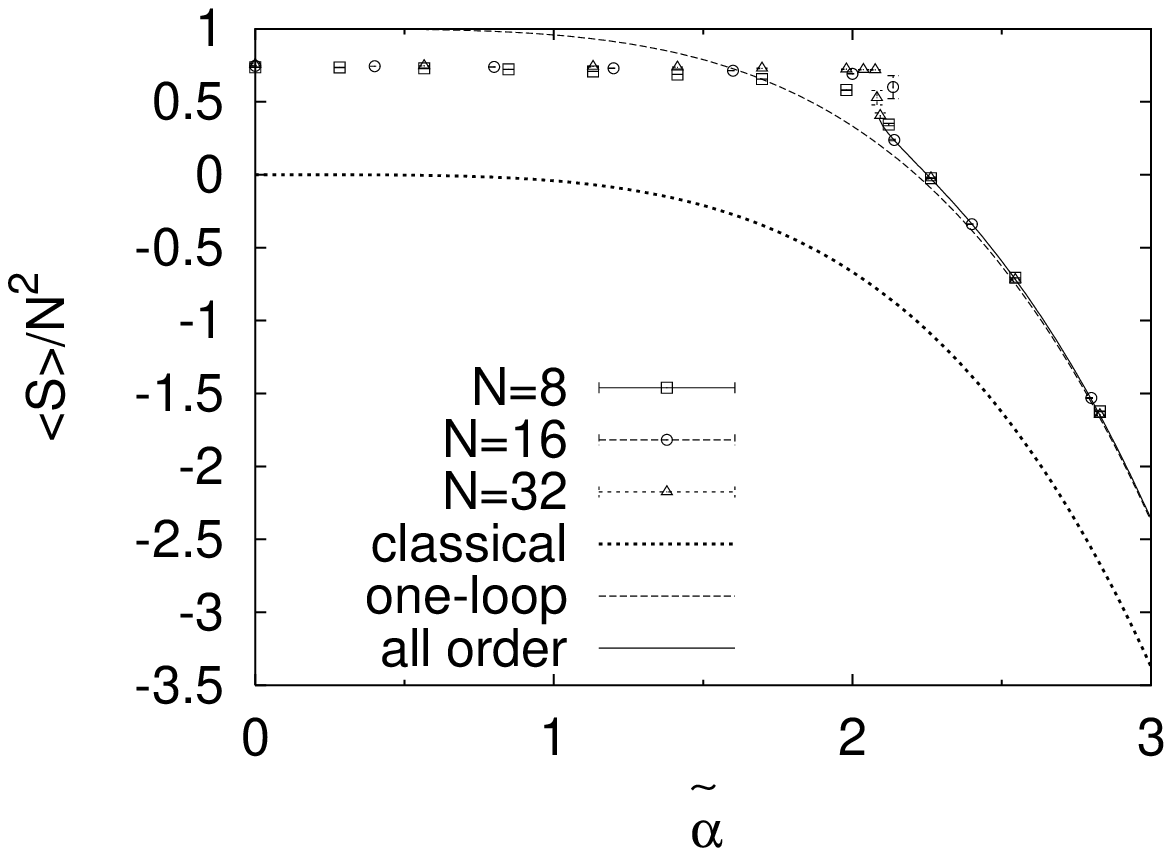,width=7.4cm}
          \epsfig{file=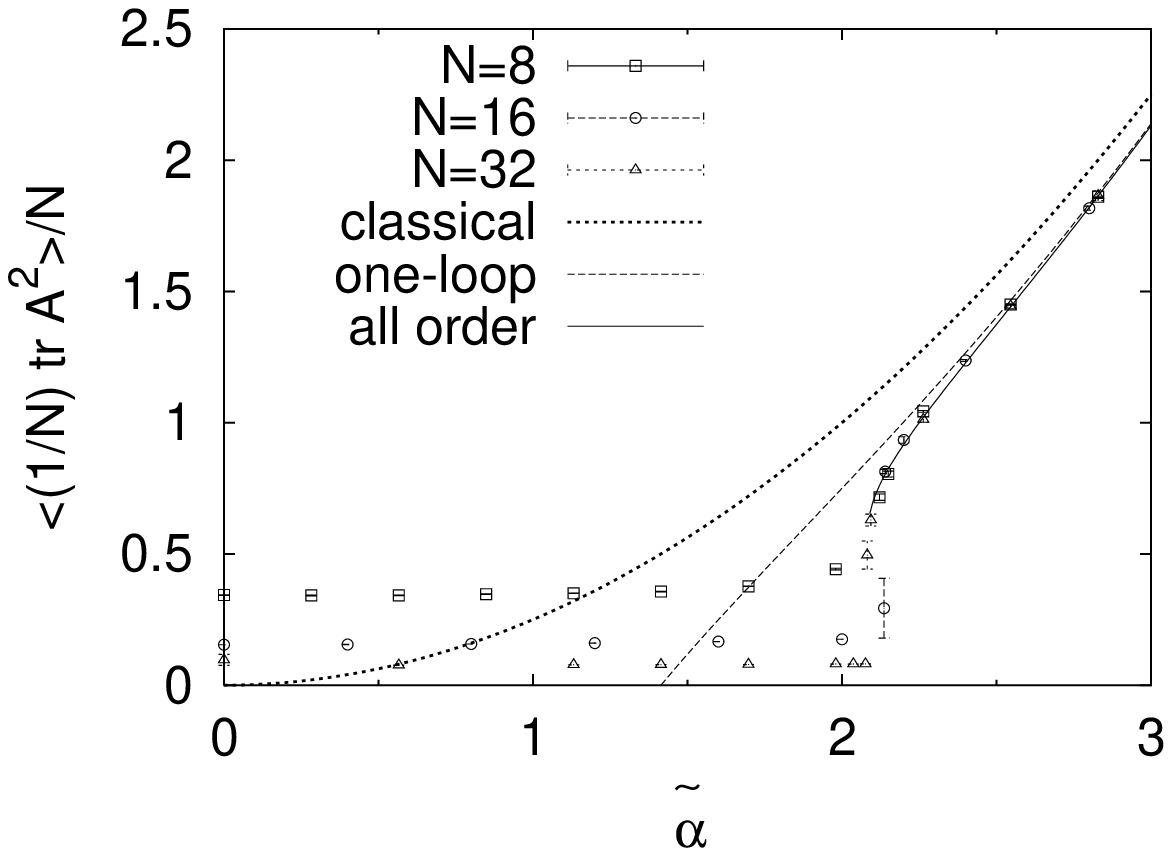,width=7.4cm}
          \epsfig{file=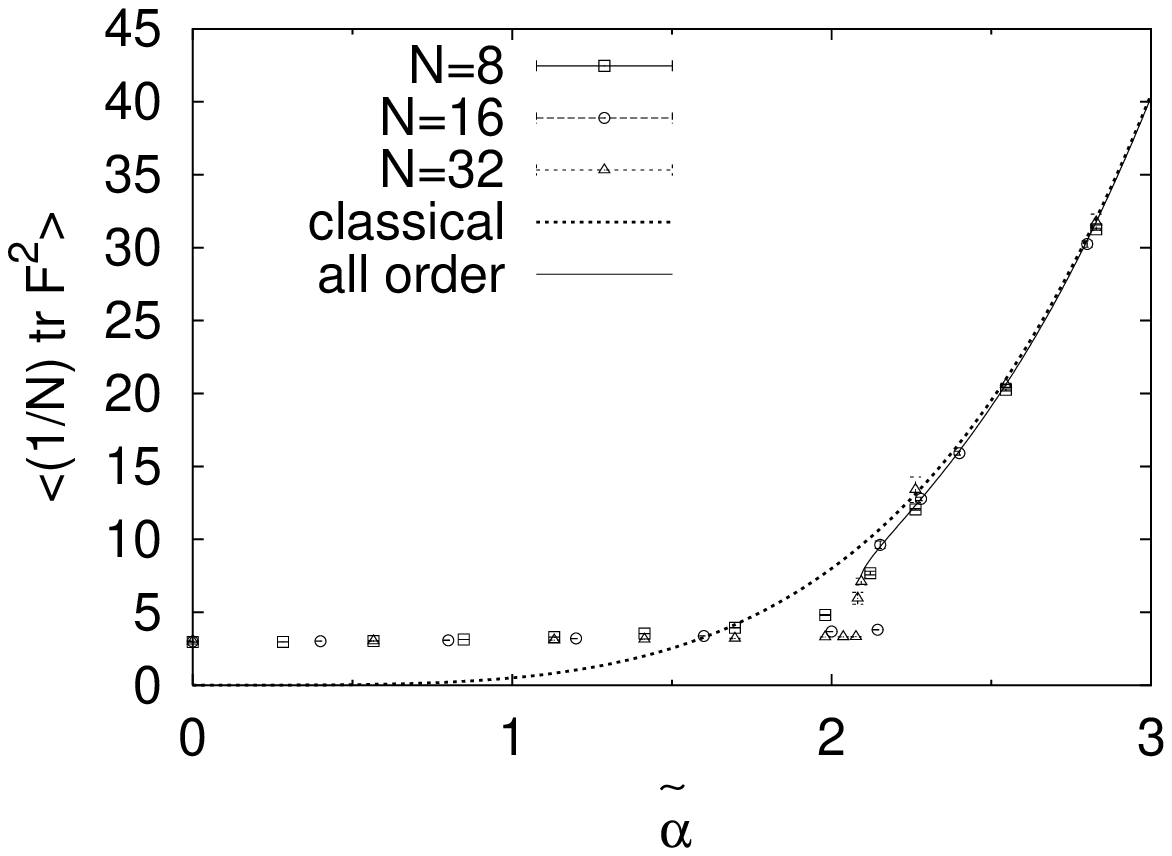,width=7.4cm}
          \epsfig{file=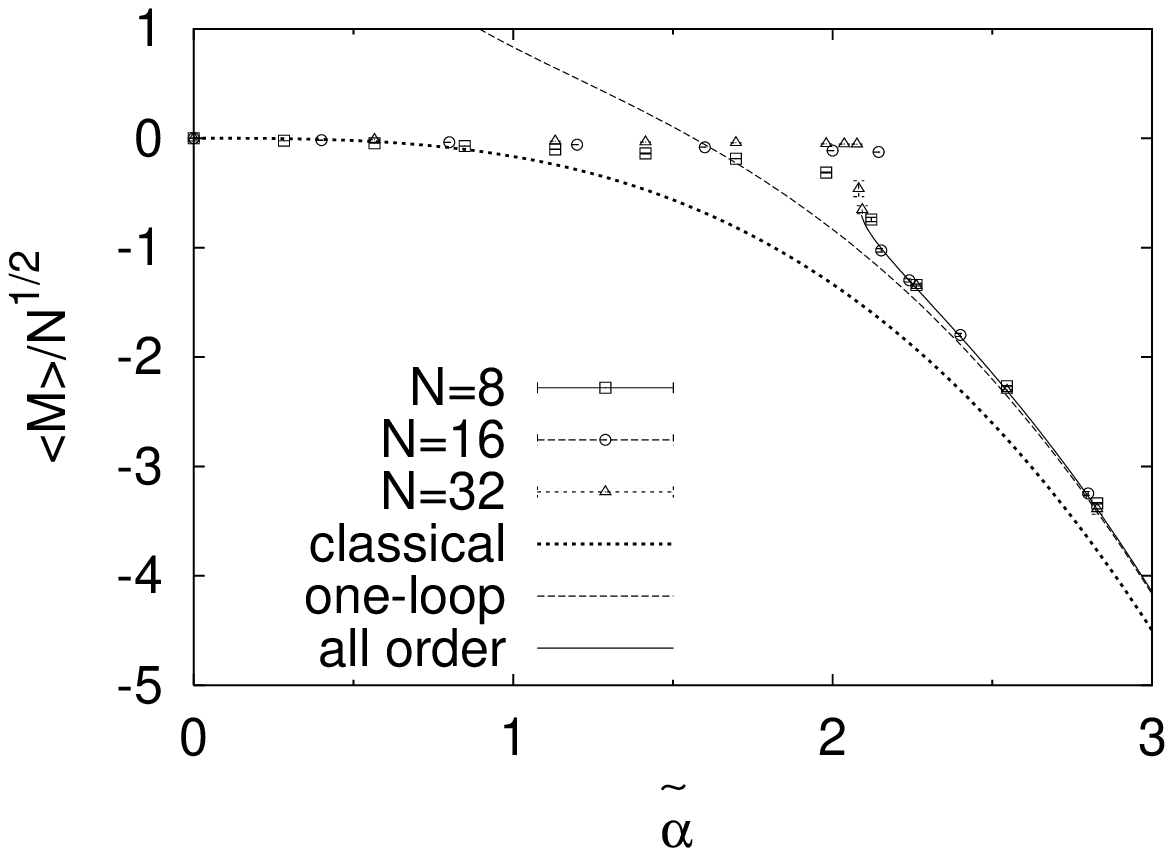,width=7.4cm}
\caption{
%% $\frac{1}{N^{2}} \langle S \rangle$ (upper left), $\frac{1}{N} 
%% \langle \frac{1}{N} \tr A^{2}_{\mu} \rangle$ (upper right), $\langle
%% \frac{1}{N} \tr F_{\mu \nu}^{2} \rangle$ (lower left) 
%% and $\frac{1}{\sqrt{N}}  \langle M \rangle$ (lower right) 
Monte Carlo results for various quantities are plotted
against ${\tilde \alpha}$ for $N=8,16,32$. 
The dotted (dashed) line
represents the classical (one-loop) result, 
while the solid line represents the ``all order'' result.
For the observable $\langle \frac{1}{N} \tr (F_{\mu \nu})^{2}
\rangle$, the one-loop result coincides with the classical result
since there is no one-loop term; See eq.\ (\ref{exact-f-sq}).}
\label{miscFS}}

In this section we compare the all order results 
(\ref{S-all-order}), (\ref{exact-a-sq}), (\ref{exact-cs-a})
and (\ref{exact-f-sq}) obtained
in the previous section with our Monte Carlo data in 
ref.\ \cite{0401038}.
Figure \ref{miscFS} shows the 
Monte Carlo results for the four observables 
as a function of ${\tilde \alpha}$ for $N=8,16,32$,
where we also plot the classical, one-loop, and all order results.
Note that the lines representing the all order results terminate
at the critical point ${\tilde \alpha} = {\tilde \alpha}_{\rm cr}$. 
%% $\frac{1}{N^{2}} \langle S \rangle$,
%% $\frac{1}{N} \langle \frac{1}{N} \tr A^{2}_{\mu} \rangle$,
%% $\langle \frac{1}{N} \tr F_{\mu \nu}^{2} \rangle$ 
%% and $\frac{1}{\sqrt{N}}  \langle M \rangle$ 
The all order results nicely reproduce
the behavior near the critical point.
This reinforces the validity of the all order calculation.

%%%%%%

\section{The dynamical gauge group revisited} 
\label{section:gauge}

  \FIGURE{\epsfig{file=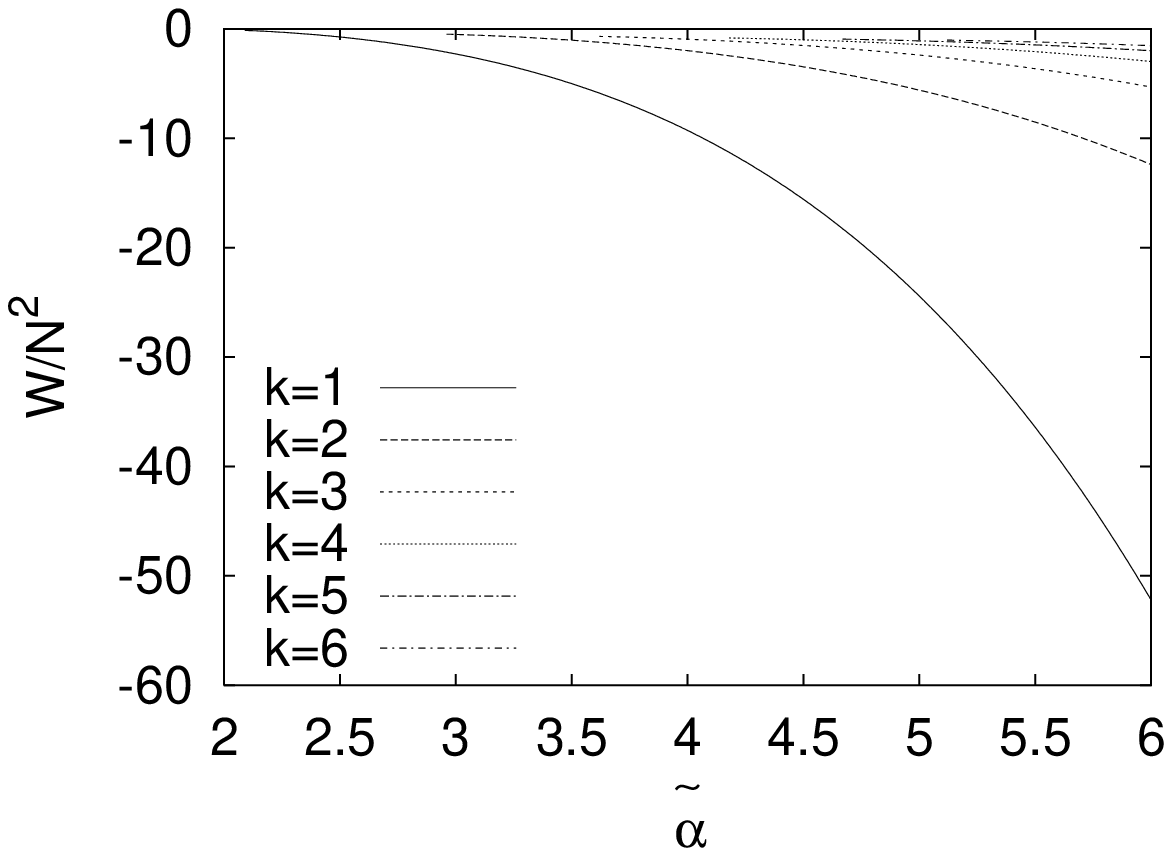,width=7.4cm}
%  \FIGURE{\epsfig{file=frees2-k.eps,width=9.0cm}
\caption{The all order result of the free energy 
for $k$ coincident fuzzy spheres 
($k=1,2,\cdots,6$) is plotted 
against ${\tilde \alpha}$.
%The lines end at the critical point
%${\tilde \alpha}_{k,{\rm cr}}$, below which the corresponding
%multi fuzzy sphere no longer exists.
%
%% The solid line, which represents the result for the single fuzzy sphere
%% ($k=1$), always gives the smallest free energy.
}
\label{frees2-k}}

The 3d YMCS model (\ref{model-def})
has various classical solutions,
which represent multi fuzzy spheres.
Among them, the solutions
 \begin{eqnarray}
  A_{\mu} = \alpha \, L^{(n)}_{\mu} \otimes {\bf 1}_{k} 
%\quad \quad \mbox{with~$N=n \, k$}
\label{k-fuzzy-sphere} 
 \end{eqnarray}
with $N=n \, k$, which correspond to $k$-coincident fuzzy spheres,
are of particular interest since they give rise to a 
noncommutative gauge theory on the fuzzy sphere with the gauge group
of rank $k$ \cite{0101102}.
If the true vacuum turns out to be described
by the solution with some $k$,
we may conclude that the gauge group of rank $k$
has been generated dynamically.
In ref.\ \cite{0401038} we discussed
this issue by comparing the one-loop effective action
evaluated {\em at the classical solutions} (\ref{k-fuzzy-sphere}).
We found that the single fuzzy sphere always has the lowest free energy
at the one-loop level in the fuzzy sphere phase.
This implies that the dynamically generated gauge group is of rank one. 
From the view point of the all order calculation discussed 
in section \ref{section:all-order-calculations},
we should consider the rescaled configuration
 \begin{eqnarray}
  A_{\mu} = \beta \, L^{(n)}_{\mu} \otimes {\bf 1}_{k} \ ,
 \label{k-fuzzy-sphere-rescaled}
 \end{eqnarray}
and evaluate the one-loop effective action {\em at its extremum},
which gives the all order result for the free energy.
In what follows we examine whether the higher order corrections
to the free energy alter the conclusion.

The one-loop effective action $\Gamma ^{(k)}$
for the rescaled configuration
(\ref{k-fuzzy-sphere-rescaled}) is given as \cite{0401038}
 \begin{eqnarray}
\lim_{N\rightarrow \infty}  
\frac{1}{N^{2}} \Gamma ^{(k)}
= \frac{1}{k^2} \left( 
\frac{1}{8} \, {\tilde \beta}^{4}
   - \frac{1}{6} \, {\tilde \alpha} \, {\tilde \beta}^{3}
 \right)    + \log {\tilde \beta} - \log k \  . 
 \label{k-eff}
 \end{eqnarray}
Similarly to the single fuzzy sphere case ($k=1$), 
the effective action (\ref{k-eff})
has a local minimum at
 \begin{eqnarray}
  {\tilde \beta} = \frac{1}{4} \, \tilde \alpha \left(
  1 +  \sqrt{1 + \delta^{(k)}} 
   +  \sqrt{ 2 - \delta^{(k)} + \frac{2}{\sqrt{1+\delta^{(k)}}}} \right) \ , 
    \label{saddle-sol-k}
 \end{eqnarray}
where
 \begin{eqnarray}
  \delta^{(k)} = 4 \, {\tilde \alpha}^{- \frac{4}{3}}
   \left[ \left( 1 + \sqrt{1 - 
   \frac{512 \, k^{2}}{27 \, {\tilde \alpha}^{4}}} \right)^{\frac{1}{3}}
        + \left( 1 - \sqrt{1 - 
   \frac{512 \, k^{2}}{27 \,  
    {\tilde \alpha}^{4}}} \right)^{\frac{1}{3}} \right] \ .
 \end{eqnarray}
The local minimum exists if and only if
${\tilde \alpha} > {\tilde \alpha}_{\rm cr}^{(k)} 
= \sqrt[4]{\frac{512 \, k^{2}}{27}}$.

In figure \ref{frees2-k} we plot the all order result for the
free energy for $k=1,2,\cdots,6$,
which is obtained by evaluating the effective action (\ref{k-eff})
at its extremum (\ref{saddle-sol-k}).
We find that the single fuzzy sphere ($k=1$) always 
has lower free energy
than the coinciding fuzzy spheres ($k \geq 2$).
Thus we conclude that
the dynamically generated gauge group is of rank one 
even if we take account of the higher order contributions.
This conclusion is consistent with our observations
in the Monte Carlo simulation \cite{0401038}.

\section{Summary} \label{section:summary}

In this paper we have clarified
some peculiar aspects of the perturbative calculation 
around fuzzy sphere solutions in matrix models at large $N$.
By direct two-loop calculation we have shown that 
the one-loop dominance, which holds for the effective action,
does not hold for observables in general.
However, we can obtain all order results for observables
from one-loop calculation by 
shifting the center of expansion to the quantum solution,
which extremizes the effective action.
The validity of this method
has been demonstrated by comparison with 
the direct two-loop calculation for the free energy
and $\langle S \rangle$.
We have also shown that the all order results for 
various observables reproduce nicely the behavior 
of our Monte Carlo data near the critical point.

From the all order results we have found that the perturbative 
expansion around the fuzzy sphere has a finite radius of convergence.
We recall that similar phenomena occur in 
exactly solvable matrix models in the planar large-$N$ limit
due to the exponential 
--- rather than factorial ---
growth of the number of planar diagrams.
In the present case the convergence property can be attributed to the 
one-loop dominance of the effective action.

The conclusions listed above should also hold 
for other matrix models, 
which incorporate higher-dimensional fuzzy manifolds.

\acknowledgments
We would like to thank Takaaki Imai, Yoshihisa Kitazawa, Yastoshi Takayama
and Dan Tomino for valuable discussions. 
K.N.\ is grateful to Denjoe O'Connor and other participants 
of the workshop {\it Fuzzy Physics '04} 
held at the Cinvestav Institute in Mexico in June 2004 
for fruitful discussions and warm hospitality.
The work of T.A.\ and J.N.\ is supported in part by Grant-in-Aid for 
Scientific Research (Nos.\ 03740 and 14740163, respectively)
from the Ministry of Education, Culture, Sports, Science and Technology.

\appendix

 \section{Details of the two-loop calculation}
 \label{section:diagram_calculation}

In this section we present the details of the two-loop
calculation. In particular we explain how eqs.\ (\ref{1PIw2})
and (\ref{s-diag-h}) are obtained.

Let us note first 
that the kinetic term (\ref{kinetic}) can be written as
\beq
S_{\rm kin} = N  \, \tr\left[ \frac{1}{2} \, \tilde A_\mu ({\cal
    P}_\lambda )^2 \tilde A_\mu + \bar c \, ({\cal P}_\lambda )^2 c
  \right]  \ ,
\label{SQ_fs}  
\eeq
where ${\cal P}_{\mu}$ is an operator acting on a $N \times N$ matrix $M$ as
 \begin{eqnarray}
  {\cal P}_{\mu}  M = [X_{\mu}, M] 
 \end{eqnarray}
with $X_{\mu}$ being the fuzzy sphere solution (\ref{fuzzy-sphere}).
The operator $({\cal P}_{\lambda})^{2}$ can be diagonalized
by the so-called ``matrix spherical harmonics''
$Y_{lm}$ ($0 \le l \le N-1$, $-l \le m \le l$), 
which form a complete basis in the
space of $N \times N$ matrices satisfying
\beqa
\frac{1}{N} \, \tr \left(  Y_{lm}^\dag  Y_{l'm'}  \right)
&=& \delta_{ll'} \, \delta_{mm'} \ ,
\label{Yortho} \\
Y_{lm}^\dag &=& (-1)^m \, Y_{l,-m} \ .
\label{Yconjg}
\eeqa
Similarly to the usual spherical harmonics, they also
possess the properties such as
   \begin{eqnarray}
  & & [L_{+}, Y_{lm}] = \sqrt{(l-m)(l+m+1)} \, Y_{l,m+1} \ ,  \nonumber \\
  & & [L_{-}, Y_{lm}] = \sqrt{(l+m)(l-m+1)} \, Y_{l,m-1} \ , \nonumber \\
  & & [L_{3}, Y_{lm}] = m \, Y_{lm} \ , \label{age-sage}
 \end{eqnarray}
where $L_{\pm} = L_{1} \pm i L_{2}$.
From these properties we find that
$Y_{lm}$ is an eigenvector of the operator 
$({\cal P}_{\lambda})^{2}$, namely
  \begin{eqnarray}
   ({\cal P}_{\lambda})^{2} \, 
   Y_{lm} = \alpha^{2} \, l \, (l+1) \, Y_{lm} \ .
  \end{eqnarray}
By expanding the matrices $\tilde{A}_\mu$, $c$ and $\bar{c}$ as
  \begin{eqnarray}
   \A_{\mu} = \sum_{l=1}^{N-1} \sum_{m=-l}^{l} \A_{\mu lm} Y_{lm} \ , 
   \hspace{2mm}
   c  = \sum_{l=1}^{N-1} \sum_{m=-l}^{l} c_{lm} Y_{lm} \ , \hspace{2mm}
   {\bar c} = \sum_{l=1}^{N-1} \sum_{m=-l}^{l} {\bar c}_{lm} Y_{lm} \ ,
  \end{eqnarray}
the propagators can be brought into the diagonal form
   \begin{eqnarray}
   \langle \A_{\mu l m} \A_{\nu l' m'} \rangle_0
 &=& \frac{1}{N \alpha^{2}} \frac{(-1)^{m}}{l \, (l+1)}
    \delta_{\mu \nu} \delta_{l, l'} \delta_{m,-m'} \ ,
   \label{propaa} \\
   \langle c_{l m} {\bar c}_{l' m'} \rangle_0
 &=& \frac{1}{N \alpha^{2}} \frac{(-1)^{m}}{l \, (l+1)}
    \delta_{l, l'} \delta_{m, -m'} \ ,
 \label{propcb}
  \end{eqnarray}
%%   \begin{eqnarray}
%%    \A_{\mu} &=& \sum_{l=1}^{N-1} \sum_{m=-l}^{l} \A_{\mu lm} Y_{lm} \ , \\
%%    c  &=&  \sum_{l=1}^{N-1} \sum_{m=-l}^{l} c_{lm} Y_{lm} \ , \\
%%    {\bar c} &=& \sum_{l=1}^{N-1} \sum_{m=-l}^{l} {\bar c}_{lm} Y_{jm} \ ,
%%   \end{eqnarray}
where the symbol $\langle \ \cdot \ \rangle_0$ represents
a VEV using only the kinetic term $S_{{\rm kin}}$
in eq.(\ref{tot-action}).

In the two-loop calculation, we use the identity
 \begin{eqnarray}
&~& \tr \, \Bigl(Y_{l_{1} m_{1}} Y_{l_{2} m_{2}} Y_{l_{3} m_{3}}
 \Bigr) \nonumber \\
& =& (-1)^{N-1} \sqrt{(2l_{1}+1)(2l_{2}+1)(2l_{3}+1)}
   \left( \begin{array}{ccc} l_{1} & l_{2} & l_{3} \\ m_{1} & m_{2} & m_{3}
 \end{array} \right) 
    \left\{ \begin{array}{ccc} l_{1} & l_{2} & l_{3} \\ L & L & L \end{array}
 \right\} \ , 
\label{3Y_identity}
 \end{eqnarray}
where $L$ is defined by $N = 2L+1$,
and Wigner's $(3j)$ and $\{ 6j \}$ symbols are given explicitly 
\cite{edmonds} by Racah's formula as
 \begin{eqnarray}
 \left( \begin{array}{ccc} l_{1} & l_{2} & l_{3} \\ m_{1} & m_{2} & m_{3}
 \end{array} \right) 
 &=& (-1)^{l_{1}-l_{2}-m_{3}} \sqrt{\Delta(l_{1}l_{2} l_{3})}
 \nonumber \\ 
 &~& \times \sqrt{(l_{1}-m_{1})! \,  (l_{1} + m_{1})! \,  (l_{2}-m_{2})! \,  (l_{2}+m_{2})! \, 
 (l_{3}-m_{3})! \,  (l_{3}+m_{3})! } \nonumber \\
 &~& \times \sum_{t} (-1)^{t} \Bigl\{ t! \,  (t-l_{2}+l_{3}+m_{1})! \, 
 (t-l_{1}-m_{2}+l_{3})! \,  (l_{1}+l_{2}-l_{3}-t)! \nonumber \\
 & & \times  (l_{1}-m_{1}-t)! \,  (l_{2}+m_{2}-t)! \Bigr\}^{-1}. 
\label{3j-symbol} \\
   \left\{ \begin{array}{ccc} l_{1} & l_{2} & l_{3} 
   \\ m_{1} & m_{2} & m_{3} \end{array} \right\} 
 &=& \sqrt{\Delta(l_{1} l_{2} l_{3}) \, \Delta(l_{1} m_{2} m_{3}) \, 
       \Delta(m_{1} l_{2} m_{3}) \, \Delta(m_{1} m_{2} l_{3})} \nonumber \\
 &~& \times \sum_{t}  (-1)^{t} (t+1)! \,  \Bigl\{ 
  (t-l_{1}-l_{2}-l_{3})! \,  (t-l_{1}-m_{2}-m_{3})! 
  \nonumber \\
 &~& \times  (t-m_{1}-l_{2}-m_{3})! \, 
  (t-m_{1}-m_{2}-l_{3})! \Bigr\}^{-1} \nonumber \\
 &~& \times 
   \Bigl\{ (l_{1}+l_{2}+m_{1}+m_{2}-t)! \,  (l_{2}+l_{3}+m_{2}+m_{3}-t)! \, 
    \nonumber \\
 &~&  \times 
    (l_{3}+l_{1}+m_{3}+m_{1}-t)! \Bigr\}^{-1}  \ ,
 \label{6j-symbol}
 \end{eqnarray}
 where 
 \begin{eqnarray}
  \Delta(abc) = \frac{(a+b-c)! \, (b+c-a)!\, (c+a-b)!}{(a+b+c+1)!} \ .
 \end{eqnarray}
 The sum of $t$ is taken over all positive integers such that no
 factorial has a negative argument. If we have a negative argument 
 in the factorial elsewhere, the $(3j)$ and $\{ 6j \}$ 
 symbols are defined to be zero. From (\ref{3Y_identity}) 
 we also obtain the formula
 \begin{eqnarray}
    [Y_{l_{1},m_{1}}, Y_{l_{2},m_{2}}] &=& \sum_{l_{3}=1}^{N-1} 
\sum_{m_{3}=-l_{3} }^{l_{3}} f^{l_{3},m_{3}}_{l_{1},m_{1},l_{2},m_{2}} 
 Y_{l_{3},m_{3}} \ , \label{commutator} \\
  f^{l_{3},m_{3}}_{l_{1},m_{1},l_{2},m_{2}} &=& (-1)^{m_{3}} (-1)^{N-1}
 \{ 1 - (-1)^{l_{1}+l_{2}+l_{3}} \} \sqrt{(2l_{1}+1)(2l_{2}+1)(2l_{3}+1)}
 \nonumber \\
 &~& \times \left( \begin{array}{ccc} l_{1} & l_{2} & l_{3}
  \\ m_{1} & m_{2} & -m_{3} \end{array} \right)
 \left\{ \begin{array}{ccc} l_{1} & l_{2} & l_{3} \\ L & L & L 
\end{array} \right\} \ .  \label{structure-y}
 \end{eqnarray}

With the help of these formulae, the contribution
of the 1PI two-loop diagrams (a)$\sim$(d)
to the coefficient $w_2(N)$ can be calculated 
as \cite{0303120}
 \begin{eqnarray}
  w_2^{\rm (a)}(N) &=&  - \frac{3}{ N} F_{1} (N) \ ,  
  \label{s-diag-a} \\
  w_2^{\rm (b)}(N)
  &=&   \frac{2}{ N} \Bigl\{ F_{1}(N) - F_{2}(N) \Bigr\} \ , 
  \label{s-diag-b} \\
  w_2^{\rm (c)}(N)
  &=& \frac{4}{ N} F_{3}(N) \ , 
  \label{s-diag-c} \\
  w_2^{\rm (d)}(N)
  &=& \frac{1}{ N} F_{1}(N) \ ,
  \label{s-diag-d}
 \end{eqnarray}
where the functions $F_{1}(N)$,
$F_{2}(N)$  and $F_{3}(N)$ are given explicitly as
\begin{eqnarray}
  F_{1}(N) &=& \frac{1}{N} \sum_{l_{1},l_{2}=1}^{N-1} \frac{(2l_{1}+1)
 (2l_{2}+1)}{l_{1} (l_{1}+1) l_{2} (l_{2}+1)} \nonumber \\
 &~&
 - \sum_{l_{1},l_{2}=1}^{N-1} (-1)^{l_{1}+l_{2}} 
\frac{(2l_{1}+1)(2l_{2}+1)}{
  l_{1}(l_{1}+1)l_{2} (l_{2}+1)} \left\{ \begin{array}{ccc}
 L & L & l_{2} \\ L & L & l_{1} \end{array} \right\} \ , 
  \label{func-f1} \\
 F_{2}(N) &=&
 \sum_{l_{1},l_{2},l_{3}=1}^{N-1} \sum_{m_{1},m_{2},m_{3}}
  \frac{(2l_{1}+1)(2l_{2}+1)(2l_{3}+1)}{l_{1}(l_{1}+1)l_{2}(l_{2}+1)
  l_{3}(l_{3}+1)} \{ 1- (-1)^{l_{1}+l_{2}+l_{3}}\}  
  \left\{ \begin{array}{ccc}
 l_{1} & l_{2} & l_{3} \\ L & L & L \end{array} \right\}^{2} \nonumber \\
  &\times& 
 \left[ 
%\prod_{i=1}^2  \sqrt{(l_{i}-m_{i})(l_{i}+m_{i}+1)}
 \frac{1}{2} \sqrt{(l_{1}-m_{1})(l_{1}+m_{1}+1)(l_{2}+m_{2})
  (l_{2}-m_{2}+1)} \right. \nonumber \\
 & & \hspace{5mm} \times  
 \left( \begin{array}{ccc} l_{1} & l_{2} & l_{3} \\
  m_{1}+1 & m_{2} & m_{3} \end{array} \right)
  \left( \begin{array}{ccc} l_{1} & l_{2} & l_{3} \\ m_{1} & m_{2}+1
 & m_{3} \end{array} \right) \nonumber \\
 & & +  
%  \prod_{i=1}^2  \sqrt{(l_{i}+m_{i})(l_{i}-m_{i}+1)}
 \frac{1}{2}
 \sqrt{(l_{1}+m_{1})(l_{1}-m_{1}+1)(l_{2}-m_{2})(l_{2}+m_{2} +1)} 
 \nonumber \\
 & & \hspace{5mm} \times 
\left( \begin{array}{ccc} l_{1} & l_{2} & l_{3} \\
  m_{1}-1 & m_{2} & m_{3} \end{array} \right)
  \left( \begin{array}{ccc} l_{1} & l_{2} & l_{3} \\ m_{1} & m_{2}-1
 & m_{3} \end{array} \right) \nonumber \\
 & & + \left. 
  m_{1} m_{2} \left( \begin{array}{ccc} l_{1} & l_{2} & l_{3} \\ m_{1} &
  m_{2} & m_{3} \end{array} \right)^{2} \right] \ , \label{func-f2} \\
 F_{3}(N) &=& \sum_{l_{1},l_{2},l_{3}=1}^{N-1} 
 \Bigl\{ 1-(-1)^{l_{1}+l_{2}+l_{3}} \Bigr\}
 \frac{(2l_{1}+1)(2l_{2}+1)(2l_{3}+1)}{l_{1} (l_{1}+1) l_{2} (l_{2}+1)
 l_{3} (l_{3}+1)} \left\{ \begin{array}{ccc} l_{1} & l_{2} & l_{3} \\
 L & L & L \end{array} \right\}^{2}. \label{func-f3}
 \end{eqnarray}
 Summing up the contributions (\ref{s-diag-a})$\sim$(\ref{s-diag-d}) 
 and using the identity $F_1(N)=-2F_2(N)$, we obtain eq.\ (\ref{1PIw2}).

 Now let us calculate the contributions from the 1PR diagrams (f)$\sim$(h).
 We first evaluate the tadpole obtained by 
 contracting two $\tilde{A}$'s in the second term in (\ref{interaction}).
 It is given as
  \begin{eqnarray}
 T_{A} &=& - N \alpha \left\{ \tr \, \wick{1}{([\A_{\mu},
  <1\A_{\nu}] [L_{\mu}, >1\A_{\nu}] ) }
  + \tr \, \wick{1}{( [<1\A_{\mu}, \A_{\nu}][L_{\mu}, >1\A_{\nu}])}
  + \tr \, \wick{1}{( [<1\A_{\mu}, >1\A_{\nu}][L_{\mu}, \A_{\nu}])} \right\}
  \nonumber \\
  &=& - N \alpha \sum_{l_{1},l_{2},l_{3}=1}^{N-1} \sum_{m_{1},m_{2},m_{3}}
  \tr \, ([Y_{l_{1} m_{1}}, Y_{l_{2} m_{2}}][L_{\mu},
     Y_{l_{3} m_{3}}]) \nonumber \\
 & & \times  
  ( \wick{11}{ \A_{\mu l_{1} m_{1}} <1\A_{\nu l_{2} m_{2}} >1\A_{\nu
    l_{3} m_{3}} - \A_{\nu l_{1} m_{1}} <2\A_{\mu l_{2} m_{2}}
    >2\A_{\nu l_{3} m_{3}} } + \wick{1}{ <1\A_{\mu l_{1} m_{1}} >1\A_{\nu 
    l_{2} m_{2}} \A_{\nu l_{3} m_{3}} } ) \nonumber \\
  &=& - \frac{1}{\alpha} \sum_{l_{1},l_{2}=1}^{N-1} \sum_{m_{1},m_{2}}
   \A_{\mu l_{1} m_{1}} \frac{(-1)^{m_{2}}}{
  l_{2} (l_{2}+1)} \nonumber \\
  & & \times \{ 2 \, \tr \, ([Y_{l_{1} m_{1}}, Y_{l_{2} m_{2}}]
  [L_{\mu}, Y_{l_{2} -m_{2}}]) + 
  \tr \, ([Y_{l_{2} m_{2}}, Y_{l_{2} -m_{2}}][L_{\mu}, Y_{l_{1} m_{1}}]) \} \ .
  \label{cubic-tadpole}
  \end{eqnarray}
 The second term of (\ref{cubic-tadpole}) vanishes since
 \begin{eqnarray}
  0 &=& \tr \, ( [L_{\mu}, Y_{l_{1} m_{1}} [Y_{l_{2} m_{2}}, 
  Y_{l_{2} -m_{2}}]] ) \nonumber \\
    &=& \tr \, ( [L_{\mu}, Y_{l_{1} m_{1}}][Y_{l_{2} m_{2}}, 
     Y_{l_{2} -m_{2}}]) \nonumber \\
    &~&+  \tr \, ( Y_{l_{1} m_{1}} [[ L_{\mu}, Y_{l_{2} m_{2}}],
     Y_{l_{2} -m_{2}}] )
     +  \tr \, ( Y_{l_{1} m_{1}} [Y_{l_{2} m_{2}}, [L_{\mu}, 
     Y_{l_{2} -m_{2}}]] ) \nonumber \\
    &=& \tr \, ( [L_{\mu}, Y_{l_{1} m_{1}}][Y_{l_{2} m_{2}}, 
     Y_{l_{2} -m_{2}}]) \nonumber \\
    &~& +  \tr \, ( Y_{l_{1} m_{1}} [[ L_{\mu}, Y_{l_{2} m_{2}}],
     Y_{l_{2} -m_{2}}] )
     -  \tr \, ( Y_{l_{1} m_{1}} [[L_{\mu}, 
     Y_{l_{2} -m_{2}}], Y_{l_{2} m_{2}}] ) \ ,
    \label{cc-tad-2}
 \end{eqnarray} 
 where the second and third terms in the last line cancel each other.
Similarly we evaluate the tadpole obtained by contracting $c$ and $\bar{c}$
in the fourth term in (\ref{interaction}). It is given as
  \begin{eqnarray}
   T_{\rm gh} &=&  - \alpha N \sum_{l_{1},l_{2},l_{3}=1}^{N-1}
 \sum_{m_{1},m_{2},m_{3}} \wick{1}{<1{\bar c}_{l_{2} m_{2}} 
   \A_{\mu l_{1} m_{1}} >1c_{l_{3} m_{3}}}
   \tr \, ( [L_{\mu}, Y_{l_{2} m_{2}}] [Y_{l_{1} m_{1}}, Y_{l_{3} m_{3}}] ) 
   \nonumber \\
   &=&  \frac{1}{\alpha} \sum_{l_{1},l_{2}=1}^{N-1}
  \sum_{m_{1},m_{2}} \A_{\mu l_{1} m_{1}}
   \frac{(-1)^{m_{2}}}{l_{2} (l_{2}+1)} 
   \tr \, ( [L_{\mu}, Y_{l_{2} -m_{2}}] [Y_{l_{1} m_{1}}, Y_{l_{2}
    m_{2}}]])  \ .
  \label{ghost-tadpole}
  \end{eqnarray}
 Thus we find that
  \begin{eqnarray}
   T_{\rm gh} = - \frac{1}{2} \,  T_{A} \ .
  \label{TTrel}
  \end{eqnarray}

  By contracting two $\tilde{A}$'s in the product of two tadpoles,
  we obtain the contribution from the diagrams (f)$\sim$(h) to the 
  coefficient $w_2(N)$, which we denote as $w_2^{\rm (f)}(N)$,
  $w_2^{\rm (g)}(N)$ and $w_2^{\rm (h)}(N)$, respectively.
 We obtain, for instance,
 \begin{eqnarray}
 w_2^{\rm (f)}(N) 
 &=& 2\alpha^{2} \sum_{l_{1},l_{2},l_{3},l_{4}=1}^{N-1}
  \sum_{m_{1},m_{2},m_{3},m_{4}} \wick{1}{<1\A_{\mu l_{1} m_{1}}
  >1\A_{\nu l_{4} m_{4}}} \frac{(-1)^{m_{2} + m_{3}}}{l_{2} (l_{2}+1)
  l_{3} (l_{3}+1)} \nonumber \\
 & &  \tr \, ([Y_{l_{1} m_{1}}, Y_{l_{2} m_{2}}]
  [L_{\mu}, Y_{l_{2} -m_{2}}]) \, 
   \tr \, ([Y_{l_{4} m_{4}}, Y_{l_{3} m_{3}}]
  [L_{\nu}, Y_{l_{3} -m_{3}}])  \nonumber \\
 &=& \frac{2}{N} \sum_{l_{1}, l_{2}, l_{3}=1}^{N-1}
 \sum_{m_{1}, m_{2}, m_{3}} \frac{(-1)^{m_{1} + m_{2} + m_{3}}}{l_{1}
 (l_{1} + 1) l_{2} (l_{2}+1) l_{3} (l_{3}+1)} \nonumber \\
 & &  \tr \, ([Y_{l_{1} m_{1}}, Y_{l_{2} m_{2}}]
  [L_{\mu}, Y_{l_{2} -m_{2}}]) \,
  \tr \, ([Y_{l_{1} -m_{1}}, Y_{l_{3} m_{3}}]
  [L_{\mu}, Y_{l_{3} -m_{3}}]) \nonumber \\ 
 &=&  \frac{2}{ N} F_{4} (N)  \ ,
 \label{cubic-cubic-result}
  \end{eqnarray}
where we have defined 
 \begin{eqnarray}
 F_{4} (N) &=& 
 \sum_{l_{1},l_{2},l_{3}=1}^{N-1}
  \sum_{m_{1},m_{2},m_{3}} \frac{(2l_{1}+1)(2l_{2}+1)(2l_{3}+1) 
  (-1)^{m_{1} + m_{2} + m_{3}}}{l_{1} (l_{1}+1) l_{2} (l_{2}+1) l_{3}
  (l_{3}+1)} \nonumber \\
 &~&\times (2 - 2(-1)^{l_{1}})\left\{ \begin{array}{ccc} l_{1} & l_{2} & l_{2}
 \\ L & L & L \end{array} \right\} \left\{ \begin{array}{ccc} l_{1}
  & l_{3} & l_{3} \\ L & L & L \end{array} \right\} 
 \nonumber \\
 &~& \times \left[ 
    m_{2} m_{3} \left( \begin{array}{ccc} l_{1}
  & l_{2} & l_{2} \\ m_{1} & m_{2} & -m_{2} \end{array} \right)
    \left( \begin{array}{ccc} l_{1}
  & l_{3} & l_{3} \\ -m_{1} & m_{3} & -m_{3} \end{array} \right)
   \right. \nonumber \\
 & & + \frac{1}{2}  \sqrt{(l_{2}+m_{2})(l_{2}-m_{2}+1)(l_{3}
  - m_{3})(l_{3} + m_{3} + 1)} \nonumber \\
 & & \times 
   \left( \begin{array}{ccc} l_{1}
  & l_{2} & l_{2} \\ m_{1} & m_{2} & -m_{2}+1 \end{array} \right) 
 \left( \begin{array}{ccc} l_{1}
  & l_{3} & l_{3} \\ -m_{1} & m_{3} & -m_{3}-1 \end{array} \right)
   \nonumber \\
  & & + \frac{1}{2}  \sqrt{(l_{2}-m_{2})(l_{2}+m_{2}+1)(l_{3}+m_{3})
   (l_{3}-m_{3}+1)} \nonumber \\
  & & \times \left. 
   \left( \begin{array}{ccc} l_{1}
  & l_{2} & l_{2} \\ m_{1} & m_{2} & -m_{2}-1 \end{array} \right) 
  \left( \begin{array}{ccc} l_{1}
  & l_{3} & l_{3} \\ -m_{1} & m_{3} & -m_{3}+1 \end{array} \right)
   \right].
  \label{cubic-cubic-result2} 
  \end{eqnarray}
 Due to the relation (\ref{TTrel}), the contributions from the other
 diagrams are readily obtained as
  \begin{eqnarray}
  w_2^{\rm (g)}(N) &=& 
  \frac{1}{2 N} F_{4}(N) \ ,
  \label{ghost-ghost-result2} \\
  w_2^{\rm (h)}(N) &=& 
    - \frac{2}{N} F_{4} (N) \ .
  \label{cubic-ghost-result2} 
  \end{eqnarray}
 Summing up all the contributions, we obtain eq.\ (\ref{s-diag-h}).

\end{document}